\documentclass{emulateapj}
\usepackage{natbib}
\usepackage{epsfig}
\usepackage{graphicx}
\usepackage{subfigure}
\usepackage{float}
\usepackage{amsmath}
\usepackage{color}
\usepackage{amssymb}
\usepackage{amsfonts} 
\usepackage{units} 
\usepackage{bm} 
\usepackage{multirow} 
\usepackage[colorlinks,linkcolor=blue,anchorcolor=green,citecolor=blue]{hyperref} 
\shorttitle{Tight constraint on photon mass from pulsar spindown}
\shortauthors{Yang \& Zhang} 
\begin{document} 

\title{Tight constraint on photon mass from pulsar spindown}

\author{Yuan-Pei Yang\altaffilmark{1,2} and Bing Zhang\altaffilmark{1,3,4}}

\affil{$^1$Kavli Institute for Astronomy and Astrophysics, Peking University, Beijing 100871, China;\\ 
$^2$ KIAA-CAS Fellow, yypspore@gmail.com;\\
$^3$ Department of Astronomy, School of Physics, Peking University, Beijing 100871, China \\
$^4$ Department of Physics and Astronomy, University of Nevada, Las Vegas, NV 89154, USA; zhang@physics.unlv.edu}

\begin{abstract}
Pulsars are magnetized rotating compact objects. They spin down due to magnetic dipole radiation and wind emission. If photon has a nonzero mass, the spin down rate would be smaller than the zero mass case. We show that an upper limit of the photon mass, i.e.  $m_\gamma\lesssim h/Pc^2$, may be placed if a pulsar with period $P$ is observed to spin down. Recently, a white dwarf (WD) --- M dwarf binary, AR Scorpii was discovered to emit pulsed broadband emission with pulses. The spin-down luminosity of the WD can comfortably power the non-thermal radiation from the system. Applying our results to the WD pulsar with  $P=117~\rm{s}$, we obtain a stringent upper limit of the photon mass between $m_\gamma<6.3\times10^{-50}~\rm{g}$ assuming a vacuum dipole spindown, and $m_\gamma<9.6\times10^{-50}~\rm{g}$ assuming a spindown due to a fully developed pulsar wind.
\end{abstract}

 
\section{Introduction}

Massive electrodynamics, i.e. the electrodynamics with non-zero photon mass, can be described by the de Broglie-Proca theory \citep{deb22,deb23,deb40,pro36b,pro36c,pro36d,pro36a,pro37,pro38}, which is invariant for the Lorentz-Poincar\'e transformation. The photon-mass bounds may be established by specifying the microscopic origin of the mass \citep[e.g.][]{ade07,bon17b}, and experimentally one may directly constrain the zero-mass hypothesis for photon.
Since it is impossible to fully prove that the photon rest mass is exactly zero via experiments, the only experimental strategy is to place ever tighter upper limits on photon rest mass.  
Due to the uncertainty principle, there is an ultimate upper limit of photon rest mass \citep[e.g.][]{gol71,tu05}, i.e., $m_\gamma\lesssim\hbar/c^2T\simeq 10^{-66}~{\rm g}$, where $T\simeq 10^{10}~{\rm yr}$ corresponds to the age of the universe. 
Whereas such an upper limit on a single particle is impossible to place, an ensemble of particles might produce visible effects at the classical level.

From the theoretical point of view, the photon-mass correction to Maxwell's equations would cause changes to the classical electromagnetic properties. One can then constrain the photon mass through testing these properties.
So far, the photon mass has been constrained via different methods, which can be divided into secure and speculative results \citep{gol10}. The former includes the frequency dependence of the speed of light ($m_\gamma<1.5\times10^{-47}~\rm{g}$) \citep{lov64,wu16,zha16,wei16,bon16,bon17,sha17}, dispersion in the ionosphere ($m_\gamma<10^{-46}~\rm{g}$) \citep{kro71}, Coulomb's law ($m_\gamma<2\times10^{-47}~\rm{g}$) \citep{wil71}, Jupiter's magnetic field ($m_\gamma<7\times10^{-49}~\rm{g}$) \citep{dav75}, and solar wind magnetic field ($m_\gamma<(2-3)\times10^{-51}~\rm{g}$) \citep{ryu97,ryu07,ret16}. The latter includes extended Lakes method ($m_\gamma<10^{-49}-10^{-52}~\rm{g}$) \citep{lak98,luo03b,luo03a,gol03}, Higgs mass for photon (no limit feasible) \citep{ade07}, cosmic magnetic fields ($m_\gamma<10^{-59}~\rm{g}$) \citep{yam59,chi76,ade07}, and so on. 

If the photon has nonzero mass, the dispersion relation would be 
\begin{equation}
\omega^2=c^2k^2+\mu^2c^2,
\end{equation} 
where 
\begin{equation}
\mu\equiv m_\gamma c/\hbar,
\end{equation} 
and $m_\gamma$ is the photon mass (see Appendix). This is the standard energy-momentum expression in the special theory of relativity, which is similar to the plasma dispersion relation (with the plasma frequency  $\omega_p$ replaced by $\mu c$). This dispersion relation means that photons with different energies have different velocities. 
One important requirement of the energy-momentum equation is that the frequency of the free electromagnetic wave must satisfy $\omega>\mu c$, or $\hbar \omega > m_\gamma c^2$.  
Therefore, one direct way to constrain the photon mass is to detect the electromagnetic wave at extremely low frequencies so that $m_\gamma<h\nu/c^2\simeq7\times10^{-47}~\unit{g}~(\nu/10~\unit{Hz})$, e.g. the Schumann resonances at $\nu\sim8~\unit{Hz}$ \citep[see][]{sch52,bal60,jac62}. At even lower frequencies (e.g. $\lesssim 1~\unit{Hz}$), the detection of the electromagnetic waves is very difficult. One possible method to study massive electrodynamics is through studying the modification of the radiation mechanisms at such low frequencies.
We note that for the magnetic dipole radiation, the angular frequency of the electromagnetic wave is equal to the angular frequency of rotation, i.e. $\omega\sim\Omega$. One natural question arises: what happens if $\Omega<\mu c$ for a magnetic dipole?

In this paper, we propose a new method to obtain a limit of the photon mass by applying the spin down information of pulsars. 
Since the Gauss units have been adopted in the literature to study pulsar spin down dynamics, we stick to this unit system throughout the paper.

\section{Pulsar spin down with nonzero photon mass}

Traditional pulsars are rapidly spinning magnetized neutron stars. White dwarf pulsars have been theoretical expected and recently proven to exist \citep[e.g.][]{zha05,mar16,gen16}.  The rotational kinetic energy luminosity of a pulsar is given by $\dot E=(2/5)MR^2\Omega\dot\Omega$, where $M$ is the pulsar mass, $R$ is the pulsar radius, and $\Omega$ is the angle velocity. Observations show that $\dot\Omega<0$ in general, i.e. pulsars spin down. Since gravitational wave spindown is negligible for slow-rotators such as radio pulsars, magnetars, and WD pulsars, pulsar spindown is naturally attributed to electromagnetic torques (due to magnetic dipole radiation or a wind power).

\subsection{Vacuum case: Magnetic dipole radiation}

The simplest model for pulsar spindown is magnetic dipole radiation. For $m_\gamma=0$, the energy losses of the vacuum magnetic dipole radiation energy is given by $L=B_p^2R^6\Omega^4\sin^2\theta/6c^3$, where $B_p$ is the polar magnetic field, and $\theta$ is the angle between the magnetic and rotational axes. The radiation energy originates from the rotational kinetic energy of the pulsar, causing the spindown of the pulsar.

\begin{figure}[H]
\centering
\includegraphics[angle=0,scale=0.3]{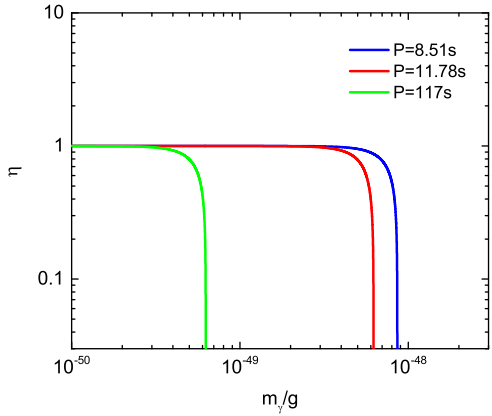}
\caption{$\eta-m_\gamma$ relation for the vacuum case. The blue, red and green lines denote $P=8.51~\rm{s}$ for PSR J2144-3933, $P=11.78~\rm{s}$ for 1ES 1841-045, and $P=117~\rm{s}$ for the WD pulsar in AR Scorpii, respectively.}\label{fig1}
\end{figure}

We calculate the magnetic dipole radiation with $m_\gamma\neq0$. Following \citet{cra84} (see Appendix), we obtain the radiation power of the magnetic dipole field in vacuum, i.e.
\begin{eqnarray}
L_m=\frac{m^2\Omega}{3}\left(\frac{\Omega^2}{c^2}-\mu^2\right)^{1/2}\left(\frac{\Omega^2}{c^2}+\frac{\mu^2}{2}\right)\label{dp0}
\end{eqnarray}
for $\Omega>\mu c$, and $L_m=0$ for $\Omega\leqslant\mu c$. Here $m$ is the magnetic dipole moment.
For $\mu=0$, the magnetic dipole radiation power reduces to the classical result, i.e.,
$L_{m,0}=(1/3)m^2\Omega^4/c^3$. We define
\begin{eqnarray} 
\eta\equiv\frac{L_m}{L_{m,0}}=\left(1-\frac{\mu^2 c^2}{\Omega^2}\right)^{1/2}\left(1+\frac{\mu^2 c^2}{2\Omega^2}\right)\label{dp}
\end{eqnarray}
for $\Omega>\mu c$, and $\eta=0$ for $\Omega\leqslant\mu c$. $\eta$ characterizes the correction of non-zero photon mass effect. The $\eta-m_\gamma$ relation for the vacuum case is presented in Figure \ref{fig1}. 
In general, the observed period of the pulsar is very accurate with error $\ll 1~\unit{s}$. Even for the WD pulsar in AR Scorpii, the relative uncertainty  is of the order $\sim 1\%$. The main uncertainty of the photon mass limit is from that of $\eta$. 
Observationally, quantifying $\eta$ needs to independently measure $L_m$ and $L_{m,0}$. Since pulsar spindown is naturally attributed to the magnetic dipole radiation in the vacuum case, one has $L_m\simeq\dot E=(2/5)MR^2\Omega\dot\Omega$, which may be derived from pulsar mass $M$, radius $R$, and spin parameters $\Omega$ and $\dot \Omega$. On the other hand, the magnetic dipole radiation without photon mass is given by $L_{m,0}\simeq\Omega^4 R^6B_p^2/6c^3$ (Here, we have adopted $\theta=\pi/2$. For the case with $\theta\neq\pi/2$, the constraint on the photon mass would be better), which requires an independent measurement of the polar cap magnetic field at surface, $B_p$. This field may be measured or constrained via other methods such as MHD pumping, Zeeman splitting\footnote{In a strong magnetic field with $B\gg B_{\rm crit}\equiv m_e^2e^3c/\hbar^3= 2.4\times10^9~\unit{G},$, the Coulomb potential is treated as a perturbation to the magnetic interaction, leading to a correction of atomic spectral lines \citep[e.g.][]{lai15}. For neutron-star pulsars with $B\gg10^9~\unit{G}$, such a correction should be considered when studying Zeeman splitting. However, for magnetized white-dwarf pulsars with $B\sim(10^6-10^9)~\unit{G}$, this strong field correction could be neglected.}, cyclotron lines, and properties of magnetar bursts. Finally, one has
\begin{eqnarray}
\eta\simeq\frac{\dot E}{L_{m,0}}=\frac{12c^3M\dot\Omega}{5B_p^2R^4\Omega^3}.\label{eta}
\end{eqnarray}
Once $\eta$ is derived from $M$, $R$, $B_p$, $\Omega$ and $\dot\Omega$ according to Eq.(\ref{eta}), one can put it in Eq.(\ref{dp}) to calculate the photon mass.

However, there is an immediate, most conservative upper limit of photon that can be readily derived. According to Eq.(\ref{dp0}) and Eq.(\ref{dp}), as long as $\eta > 0$ is satisfied, {\em which is true as long as a pulsar is observed to spin down}, one should have $\mu<\Omega/c$. A robust photon mass upper limit can be set to $m_{\gamma,{\rm crit}}$, i.e.
\begin{equation}
m_\gamma<m_{\gamma,{\rm crit}}\equiv h/Pc^2,\label{key}
\end{equation} 
which is shown as the a sharp cut off of $\eta$ in Figure \ref{fig1}.
Equation (\ref{key}) is essentially the result of the standard energy-momentum relation and does not depend on the detailed pulsar parameters other than the spin period $P$.

More generally, if $\eta$ can be constrained from Eq.(\ref{eta}), one can give a more stringent limit than that of Eq.(\ref{key}).
In Table \ref{tab1}, we list the results of $m_\gamma$ upper limits for the most conservative one ($\eta>0$) and three other assumed lower limit values, i.e. 0.1, 0.5, and 0.9. One can see that the improvement from the most conservative value even from $\eta\gtrsim0.9$ is not significant. As a result, for the vacuum case one does not need to measure $B_p$ to derive a robust constraint. As long as the pulsar is observed to spin down, i.e. $\eta > 0$, a conservative limit of $m_\gamma< m_{\gamma,{\rm crit}}$ can be placed.

Since $m_{\gamma,{\rm crit}}$ is inversely proportional to $P$, a pulsar with a longer period would give a more stringent constraint.
For neutron-star radio pulsars, the one with the longest period is PSR J2144-3933 \citep{you99} with $P=8.51~\rm{s}$, $\dot{P}=0.475\times10^{-15}~\rm{s~s^{-1}}$, and spin-down luminosity $\dot E=3.2\times10^{28}~{\rm erg~s^{-1}}~\left(I/10^{45}~{\rm g~cm^2}\right)$. 
For magnetars, the one with the longest period is 1ES 1841-045 \citep{dib08} with $P=11.78~\rm{s}$, $\dot{P}=3.93\times10^{-11}~\rm{s~s^{-1}}$, and   $\dot E=0.95\times10^{33}~{\rm erg~s^{-1}}~\left(I/10^{45}~{\rm g~cm^2}\right)$.
For the WD pulsars, some magnetized WDs \citep{wic00} with $B\sim10^6-10^9~\rm{G}$ have periods around one hour \citep{fer97}. However, no spindown measurements have been reported.
The WD --- M dwarf binary system, AR Scorpii, was recently reported to emit pulsed broadband emission \citep{mar16}. The WD pulsar in AR Scorpii has a measured spindown parameters $P=117~\rm{s}$, $\dot{P}=3.9\times10^{-13}~\rm{s~s^{-1}}$, and $\dot E=1.5\times10^{33}~{\rm erg~s^{-1}}$. The mean luminosity of AR Scorpii is $1.7\times10^{32}~{\rm erg~s^{-1}}$, which includes the thermal emission of $4.4\times10^{31}~{\rm erg~s^{-1}}$ from the stellar components.  Therefore, the non-thermal emission from AR Scorpii could be comfortably powered by the magnetic dipole radiation of the white dwarf \citep{mar16,gen16}. Such a long period white dwarf pulsar can provide a stringent constraint on the photon mass.

In Figure \ref{fig1}, the blue, red and green lines denote the periods of PSR J2144-3933, 1ES 1841-045 and the WD pulsar in AR Scorpii, respectively. 
The respective constrained upper limits of the photon mass 
are shown in Table \ref{tab1}:
for PSR J2144-3933, one has $m_\gamma<8.6\times10^{-49}~\rm{g}$;
for 1ES 1841-045, one has $m_\gamma<6.2\times10^{-49}~\rm{g}$;
and for the WD pulsar in AR Scorpii, one has 
\begin{equation}
m_\gamma<6.3\times10^{-50}~\rm{g}.
\end{equation}

\subsection{Nonvacuum case: Pulsar wind}

If the magnetic axis is parallel to the rotation axis, the magnetic dipole radiation is zero in vacuum. However, active pulsars are believed to be surrounded by a magnetosphere, from which a continuously outflow is launched from the open field line regions as a pulsar wind \citep{gol69}.
The outflowing plasma exerts an electromagnetic torque on the pulsar, so that the pulsar would spin down due to the existence of a pulsar wind \citep{harding99,xu01,con06,ton13}.

\begin{table*}{}
    \begin{center}
    \scriptsize
    \caption{The upper limits of the photon mass}
    \begin{tabular}{cccccccccccccccc}
    \hline
    \hline
    \multirow{2}{*}{Sources} & \multirow{2}{*}{Period/s}& \multicolumn{7}{c}{$m_\gamma/10^{-49}~\unit{g}$ in vacuum case} &  \multicolumn{6}{c}{$m_\gamma/10^{-49}~\unit{g}$ in nonvacuum case}\\ 
    & & $\eta>0$ & & $\eta>0.1$ & & $\eta>0.5$ & &$\eta>0.9$ & & $\eta>0.1$ & &  $\eta>0.5$ & & $\eta>0.9$\\
    \hline
    PSR J2144-3933 &$8.51$& $<m_{\gamma,{\rm crit}}=8.6$ & & $<8.5$ & & $<8.1$ & & $<5.9$ & & $<20.3$ & & $<9.3$ & & $<3.1$\\
    1ES 1841-045 &$11.78$& $<m_{\gamma,{\rm crit}}=6.2$ & & $<6.2$ & & $<5.8$ & & $<4.2$ & & $<14.7$ & & $<6.7$  & & $<2.2$\\
    WD in AR Scorpii &$117$& $<m_{\gamma,{\rm crit}}=0.63$ & & $<0.62$ & & $<0.59$ & & $<0.43$ & & $<1.48$ & & $<0.67$ & & $<0.22$\\
    \hline
    \hline
    \end{tabular}\label{tab1}
    \end{center}
\end{table*} 

To quantify the effect of non-zero photon mass on the wind spindown rate, we first calculate the magnetic dipole field with $m_\gamma\neq0$. We consider a statistic field solution (see Appendix): 
$(\nabla^2-\mu^2)\bm{A}=-4\pi\bm{J}/c$. The corresponding solution is \citep{jac62}
\begin{eqnarray}
\bm{A}&=&\frac{1}{c}\int \bm{J}(\bm{x'})\frac{e^{-\mu |\bm{x}-\bm{x'}|}}{|\bm{x}-\bm{x'}|}d^3x'\nonumber\\
&=&-\bm{m}\times\nabla\int \delta(\bm{x'})\frac{e^{-\mu |\bm{x}-\bm{x'}|}}{|\bm{x}-\bm{x'}|}d^3x'.
\end{eqnarray}
Note that $\bm{J}=c(\nabla\times\bm{\mathcal{M}})$, where $\bm{\mathcal{M}}=\bm{m}\delta(\bm{x})$ is the magnetization for the dipole approximation.
According to $\bm{B}=\nabla\times\bm{A}$, one has \citep[e.g.][]{jac62}
\begin{eqnarray}
\bm{B}&=&-\nabla\times\left(\bm{m}\times\nabla\frac{e^{-\mu r}}{r}\right)\nonumber\\
&=&[3\bm{n}(\bm{n}\cdot\bm{m})-\bm{m}]\left(1+\mu r+\frac{\mu^2 r^2}{3}\right)\frac{e^{-\mu r}}{r^3}\nonumber\\
&-&\frac{2}{3}\mu^2\bm{m}\frac{e^{-\mu r}}{r}.
\end{eqnarray}
The magnetic field components in a spherical coordinate system are given by
\begin{eqnarray}
B_r&=&2m\frac{e^{-\mu r}}{r^3}\left(1+\mu r+\frac{\mu^2r^2}{3}\right)\cos\theta-\frac{2\mu^2m}{3}\frac{e^{-\mu r}}{r}\cos\theta,\nonumber\\
B_\theta&=&m\frac{e^{-\mu r}}{r^3}\left(1+\mu r+\frac{\mu^2r^2}{3}\right)\sin\theta+\frac{2\mu^2m}{3}\frac{e^{-\mu r}}{r}\sin\theta. \label{mag}\nonumber\\
\end{eqnarray}
\begin{figure}[H]
\centering
\includegraphics[angle=0,scale=0.45]{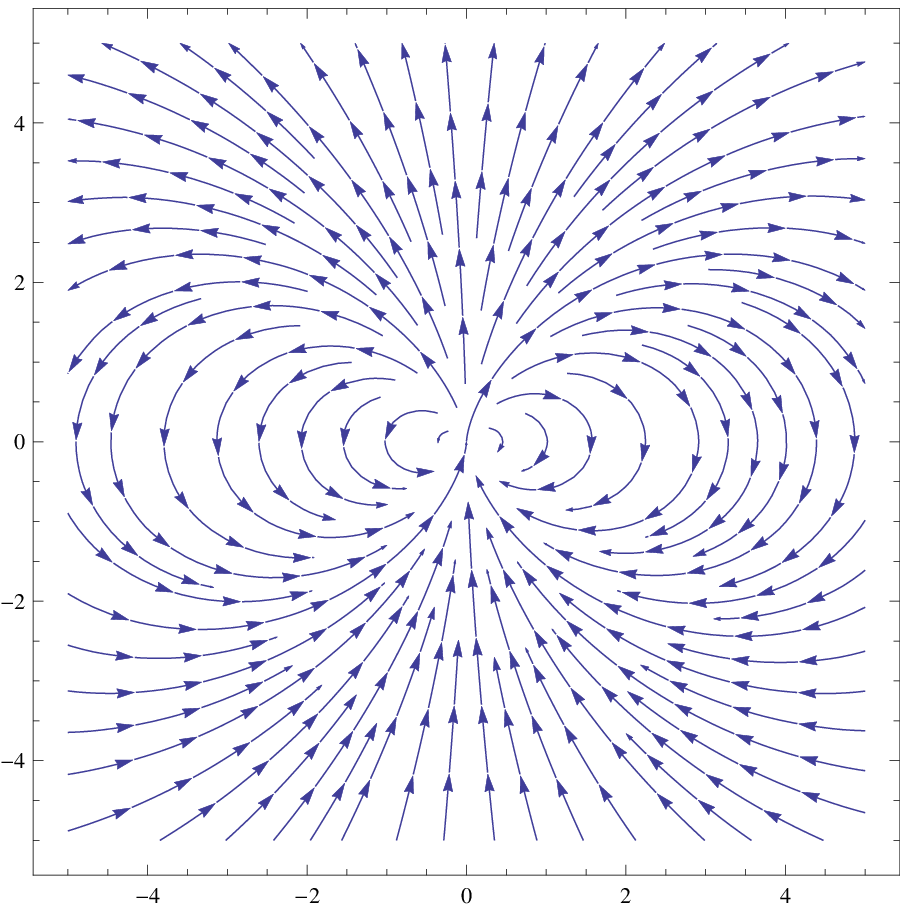}
\includegraphics[angle=0,scale=0.45]{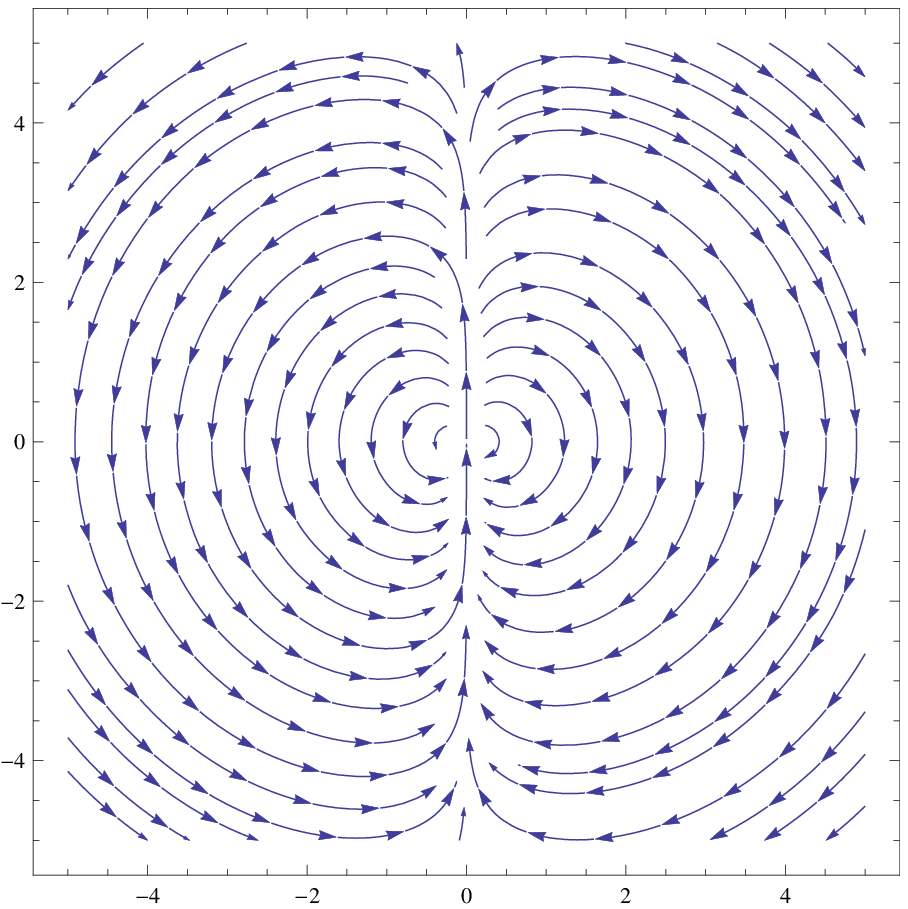}
\caption{The magnetic dipole field with $m_\gamma=0$ (left) and $m_\gamma\neq0$ (right).}\label{fig2}
\end{figure}
The field-line equation $dr/B_r=rd\theta/B_\theta$ can be written as 
$dr/d\theta=2r(1+\mu r)\cot\theta/(1+\mu r+\mu^2 r^2)$. 
Interestingly, for $\mu\rightarrow\infty$, one has $dr/d\theta\simeq2\cot\theta/\mu\rightarrow0$ for $\theta\gg0$, which means that the magnetosphere would approach a 3-dimensional sphere, as shown in Figure \ref{fig2}.
Integrating the field-line equation, one obtains
$L\sin^2\theta=r e^{\mu r}/(1+\mu r)$.
Thus, the open field line angle at the pulsar surface becomes 
$\sin\theta_{c}=\xi\left(R/R_{\rm LC}\right)^{1/2}$
where
\begin{eqnarray}
\xi&\equiv&\left(\frac{1+\mu R_{\rm LC}}{1+\mu R}\right)^{1/2}e^{\mu(R-R_{\rm LC})/2}\nonumber\\
&\simeq&\left(1+\frac{\mu c}{\Omega}\right)^{1/2}e^{-\mu c/2\Omega}
\end{eqnarray}
denotes the correction factor with respect to the zero photon mass case, and 
$R_{\rm LC}\equiv c/\Omega$ is the light cylinder radius. As shown in Figure \ref{fig2} and Figure \ref{fig3}, the larger the photon mass $m_\gamma$, the smaller the polar cap.

\begin{figure}[H]
\centering
\includegraphics[angle=0,scale=0.3]{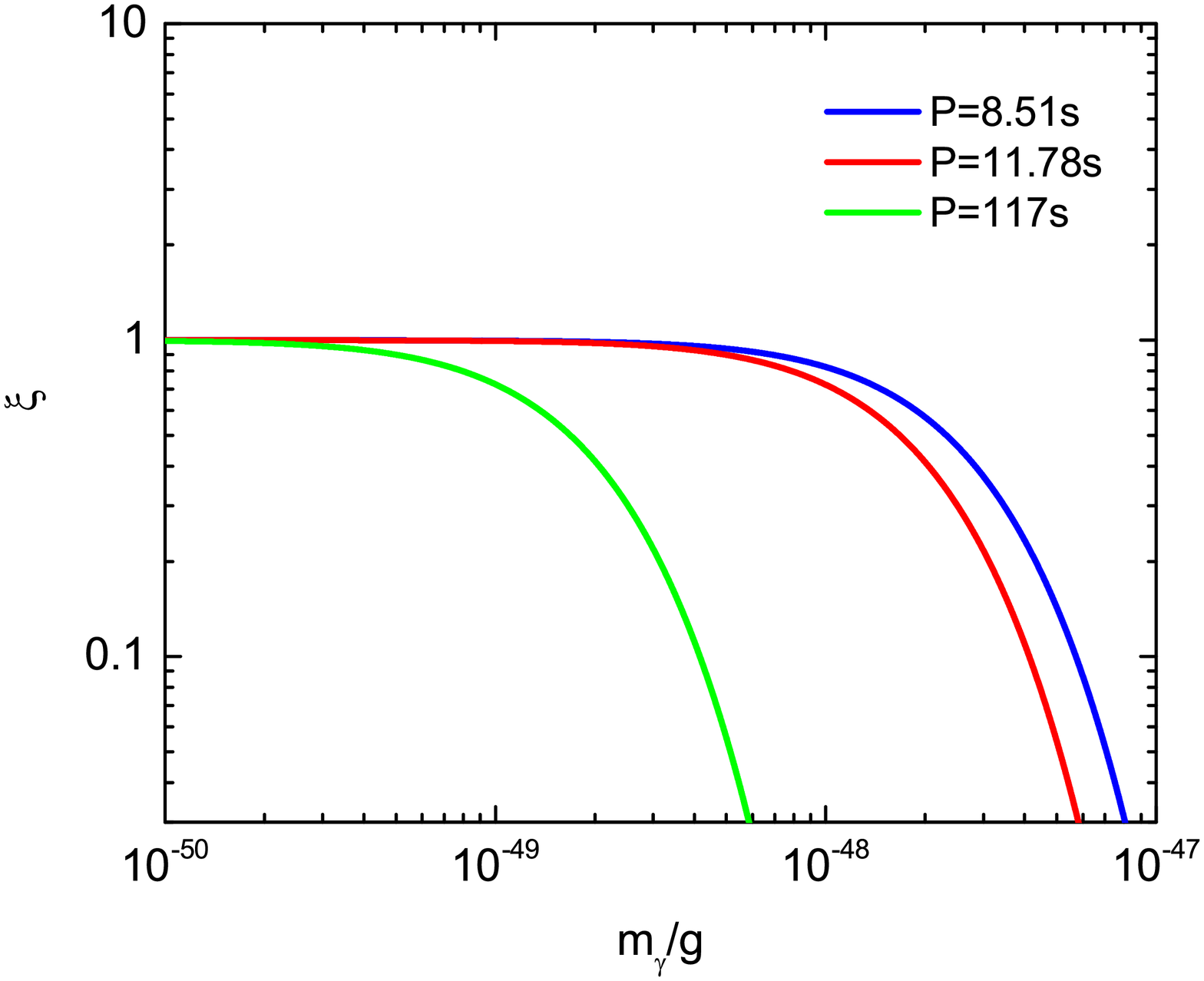} 
\caption{$\xi-m_\gamma$ relation for the non-vacuum wind spindown case. The blue, red and green lines denote $P=8.51~\rm{s}$ for PSR J2144-3933, $P=11.78~\rm{s}$ for 1ES 1841-045, and $P=117~\rm{s}$ for the WD pulsar in AR Scorpii, respectively.}\label{fig3}
\end{figure}

\begin{figure}[H]
\centering
\includegraphics[angle=0,scale=0.3]{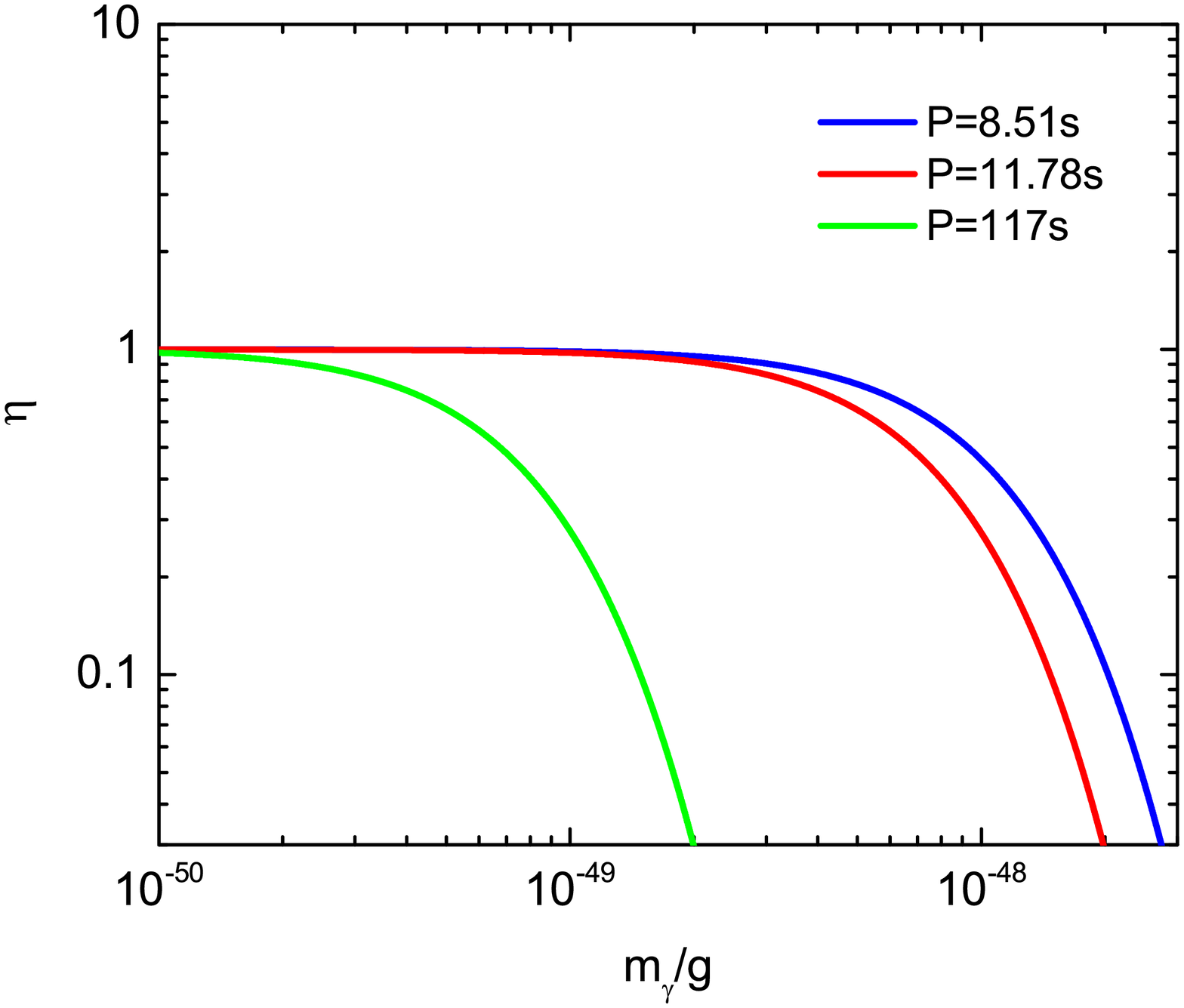}
\caption{$\eta-m_\gamma$ relation for the non-vacuum wind spindown case. The blue, red and green lines denote $P=8.51~\rm{s}$ for PSR J2144-3933, $P=11.78~\rm{s}$ for 1ES 1841-045 and $P=117~\rm{s}$ for the WD pusalr in AR Scorpii, respectively.}\label{fig4}
\end{figure}

Next, we calculate the plasma density in the magnetosphere.
Since photon mass does not affect the Lorentz force density of the matter \citep{gol71}, the Ohm's law is still
$\bm{J}=\sigma\left(\bm{E}+\bm{\upsilon}\times\bm{B}/c\right)$.
For astrophysical force-free plasmas, due to $\sigma\rightarrow\infty$ and $\bm{\upsilon}=\bm{\Omega}\times\bm{r}$, one has
$\bm{E}=-\bm{\Omega}\times\bm{r}\times\bm{B}/c$ \citep{gol69,ryu97,ryu07}.
Therefore, the charge density is given by
$\rho=(\nabla\cdot\bm{E}+\mu^2\phi)/4\pi=-\bm{\Omega}\cdot\bm{B}/2\pi c+\mu^2\phi/4\pi$ (see Appendix).
On the other hand, according to the statistic field equation:
$\rho=(-\nabla^2+\mu^2)\phi/4\pi$,
one has
$\nabla^2\phi=-4\pi\rho_{\rm GJ}$,
where
$\rho_{\rm{GJ}}\equiv-\bm{\Omega}\cdot\bm{B}/2\pi c$ is the classical Goldreich-Julian density \citep{gol69}. 
The new plasma charge density $\rho$ with $m_\gamma\neq0$ is then given by
\begin{eqnarray}
\nabla^2\left(\rho-\rho_{\rm GJ}\right)+\mu^2\rho_{\rm GJ}=0.
\end{eqnarray}
For a neutron star (white dwarf), if $m_\gamma\ll\hbar/Rc\simeq3.5\times10^{-44}~{\rm g}$ ($6.4\times10^{-47}~{\rm g}$) (established with existing photon mass limits), $R \ll R_{\rm LC}$ is satisfied. The magnetic field strength and the charge density at the pole are $B_p\simeq2m/R^3$ and $\rho\simeq\rho_{\rm GJ}$, respectively, which are consistent with the case of $m_\gamma=0$. The radius of the polar cap is
$r_p=\xi R\left(\Omega R/c\right)^{1/2}$.
The potential difference between the center and the edge of the polar cap is 
$\Delta V\simeq \Omega r_p^2B_p/2c=\xi^2\Omega^2 R^3B_p/2c^2$.
The net charged-particle flux from the polar cap is
$\dot N\simeq \pi r_p^2n_p c=\xi^2\chi\Omega^2 R^3B_p/2ec$,
where $n_p=\chi n_{_{\rm GJ}}$ with $\chi\sim1$ being the mean number density of the primary charged particles at the polar cap, which is essentially the Goldreich-Julian density.
The spin-down power of the wind is therefore approximately
\begin{eqnarray}
L_w\simeq 2e\dot N\Delta V=\xi^4\frac{\chi\Omega^4 R^6B_p^2}{2c^3}\equiv\eta L_{w,0},
\end{eqnarray}
where
\begin{eqnarray}
\eta&\equiv& L_w/L_{w,0}=\xi^4=\left(1+{\frac {\mu c}{\Omega}}\right)^2e^{-2\mu c/\Omega}\label{wp},
\label{eq:constraint2}
\end{eqnarray}
and $L_{w,0}\simeq\chi\Omega^4 R^6B_p^2/2c^3$ is the spin-down power of the wind with $m_\gamma=0$. 
This result can be also derived from the method of \citet{con99} with $\chi=1/3$ (see Appendix). Equation (\ref{eq:constraint2}), even through not an explicit expression of $m_\gamma$, can be used to derive $m_\gamma$ upper limit when a lower limit of $\eta$ is given ($m_\gamma=0$ when $\eta=1$).
The $\eta-m_\gamma$ relation for non-vacuum wind spindown case is presented in Figure \ref{fig4}. Similar to the vacuum case, the wind spin down power rapidly fall off beyond a certain $m_\gamma$ given a measured $P$.
The difference from the vacuum case is that 
there is no absolute cutoff at $m_{\gamma,{\rm crit}}=h/Pc^2$. 
In order to obtain $\eta\equiv L_w/L_{w,0}$ in Eq.(\ref{wp}), where $L_w\simeq \dot E=(2/5)MR^2\Omega\dot\Omega$ and $L_{w,0}\simeq\chi\Omega^4 R^6B_p^2/2c^3$, the measurement of the magnetic field strength is necessary (unlike the vacuum case). 

\begin{figure}[H] 
\centering
\includegraphics[angle=0,scale=0.3]{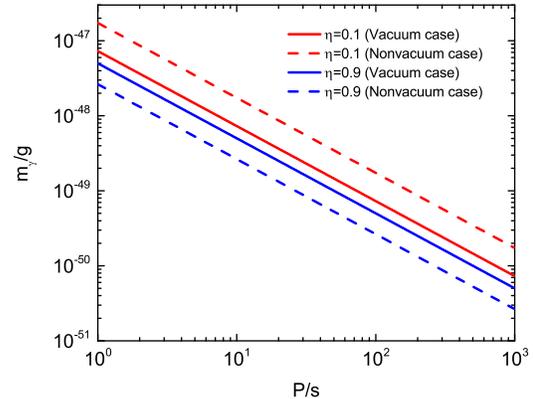}
\caption{Constraints on $m_\gamma$ for different $P$. The solid and dashed lines denote the vacuum dipole spindown case and the non-vacuum wind spindown case, respectively. We adopt $\eta=0.1~\rm{s}$ (red) and $0.9~\rm{s}$ (blue) in the calculations.}\label{fig5}
\end{figure}

Once $P$, $\dot P$ and $B_p$ are measured, one may constrain $\eta= L_w/L_{w,0}$ and
derive the upper limit of the photon mass, as shown in Figure \ref{fig5}. We still take the data from PSR J2144-3933, 1ES 1841-045 and the WD pulsar in AR Scorpii as examples. We again assume $\eta>0.1,0.5,0.9$.
The upper limits of the photon mass using different sources are shown in Table \ref{tab1}. In particular,
the WD pulsar in AR Scorpii \citep{mar16} has $P=117~\rm{s}$, $\dot{P}=3.9\times10^{-13}~\rm{s~s^{-1}}$ and $\dot E=1.5\times10^{33}~{\rm erg~s^{-1}}$. 
An independent constraint on the white dwarf surface magnetic field strength was set via the ``MHD pumping'' of the secondary star (M-dwarf) in the binary system \citep{buc16}. The magnetic field of the WD would penetrate and dissipate in the M-dwarf atmosphere and give rise to additional optical emission. The dissipation power of magnetic energy at the M-dwarf can be estimated from\footnote{In principle, the correction to magnetic energy density due to non-zero photon mass should be included. However, given the already established stringent photon mass upper limit in the literature, this correction is negligibly small. } $P_{\rm MHD}=(B_2^2/8\pi)(4\pi R_2^2\delta)(2\pi/P_{\rm b})$, where $R_2\simeq2.5\times10^{10}~\unit{cm}$ is the radius of the M dwarf star, $P_{\rm b}\simeq118~\unit{s}$ is the beat period, 
$\delta=(\eta_{\rm tur}P_{\rm b}/\pi)^{1/2}\simeq2\times10^8\eta_{\rm tur,15}^{1/2}~\unit{cm}$ is the dissipation depth of magnetic energy for the photospheric conditions of a M-type dwarf, and 
$\eta_{\rm tur} = (10^{15}~\unit{cm^2~s^{-1}})\eta_{\rm tur,15}$ is the turbulent diffusivity \citep{buc16}. If a fraction of the mean optical luminosity of AR Sco in excess of the combined stellar contributions is from MHD pumping, i.e., $P_{\rm MHD}\simeq \zeta L_+=1.3\times10^{32}\zeta~\unit{erg~s^{-1}}$, where $\zeta$ is the fraction of MHD pumping contribution, the magnetic field strength at the secondary star would be $B_2\simeq204\zeta^{1/2}\eta_{\rm tur,15}^{-1/4}~\unit{G}$. For $m_\gamma\ll\hbar/Rc\simeq6.4\times10^{-47}~{\rm g}$ (established with existing photon mass limits), the magnetic dipole field of the white dwarf would satisfy $B\propto r^{-3}$. 
The magnetic field strength at the pole of the WD would be then derived as $B_p\simeq B_2(a/R)^3\simeq6.4\times10^{8}\zeta^{1/2}\eta_{\rm tur,15}^{-1/4}M_{0.8}^{3.4}~\unit{G}$, where $a\simeq8\times10^{10}M_{0.8}^{1/3}~\unit{cm}$ is the binary separation \citep{mar16,buc16}, and $M=(0.8M_\odot)M_{0.8}$ is the white dwarf mass. According to the Hamada-Salpeter relation, the radius-mass relation of white dwarfs satisfies $R\simeq5.5\times10^8M_{0.8}^{-0.8}~\unit{cm}$. The dipole radiation luminosity with zero-mass photon is $L_{w,0}\simeq\Omega^4 R^6B_p^2/6c^3\simeq5.8\times10^{32}\zeta\eta_{\rm tur,15}^{-1/2}M_{0.8}^2~\unit{erg~s^{-1}}$. The dipole radiation luminosity with nonzero-mass photon is $L_w\simeq\dot E=(2/5)MR^2\Omega\dot\Omega\simeq 1.9\times10^{33}M_{0.8}^{-0.6}~\unit{erg~s^{-1}}$.
One therefore has $\eta\simeq 3.3\zeta^{-1}\eta_{\rm tur,15}^{1/2}M_{0.8}^{-2.6}$. Here the turbulent diffusivity is taken as $\eta_{\rm tur}\simeq(10^{14}-10^{15})~\unit{cm^2s^{-1}}$ \citep[e.g.][]{mei06,buc16} and the white dwarf mass is taken as $M\simeq(0.8-1.3)M_\odot$ \citep{mar16}. Since $\zeta<1$, one can derive $\eta>0.3$, which is the most conservative constraint based on the uncertainties of all observed quantities. As a result, for the non-vacuum wind spindown case we obtain 
\begin{equation}
m_\gamma<9.6\times10^{-50}~\unit{g}. 
\end{equation}

\section{Conclusion and Discussion}

We proposed a method of using pulsar (including neutron stars and white dwarfs) spin-down observations to set a stringent upper limit on the photon mass. In particular, for the recently observed WD pulsar in AR Scorpii, we obtained a stringent constraint on the photon mass $m_\gamma\lesssim {\rm several} \times 10^{-50}~\rm{g}$. For vacuum spindown case, the fact that the WD spin downs places a robust lower limit $m_\gamma < m_{\gamma,{\rm crit}} = h/Pc^2 = 6.3\times 10^{-50} \unit{g}$. For the non-vacuum fully developed wind spindown case, based on the constraint of surface magnetic field of the WD using MHD pumping modeling \citep{buc16}, one can derive $m_\gamma < 9.6 \times 10^{-50} \unit{g}$. In reality, the spin down behavior of the WD pulsar may be explained in the parameter regime between these two extreme models. These two derived photon mass limits can be therefore taken the bracket the true photon mass upper limit derived from the WD pulsar.

Since some magnetized white dwarfs \citep{wic00,zha05} with $B\sim10^6-10^9~\rm{G}$ have periods up to one hour and longer \citep{fer97}, we strongly urge further observations to these objects to detect their spindown behavior ($\dot P$) and to measure their magnetic field strength independently. These observations would give an upper limit around $m_\gamma\lesssim10^{-51}~\rm{g}$, which would be the most stringent limit within the secure methods. 

In our analysis, we have assumed that the magnetic field of the pulsar is dipolar. This is justified in large scale for NS and WD pulsars. Near the magnetic poles of the pulsars, multipole magnetic components may exist. For the vacuum case, even if the magnetic field is multipole, the robust limit $m_\gamma<m_{\gamma,{\rm crit}}\equiv h/Pc^2$ is still valid, since it is rooted from the standard energy-momentum relation. For the non-vacuum case, the accurate calculation of the spindown power of the multipole field is complicated. 
However, due to the contribution of the photon mass, the term of $e^{-\mu r}$ still appears in the field equation, so that the enclosed magnetic flux of the open field line would contains a factor of $e^{-\mu R_{\rm LC}}$ (see Appendix). Therefore, the spindown power of the multipole field would also be suppressed when $\Omega<\mu c$.
As a result, our derivations based on the dipole assumption would be still valid to order of magnitude.

\acknowledgments
We thank the anonymous referee and for detailed suggestions that have allowed us to improve this manuscript significantly.
This work is partially supported by the Initiative Postdocs Supporting Program (No. BX201600003), the National Basic Research Program (973 Program) of China (No. 2014CB845800) and Project funded by China Postdoctoral Science Foundation (No. 2016M600851). Y.-P.Y. is supported by a KIAA-CAS Fellowship.

\appendix

\section{Maxwell equations with nonzero photon mass}

The classical Maxwell equations and the corresponding Lagrangian are based on the hypothesis that the photon mass is zero. If photon has a non-zero mass, one can modify the Lagrangian density by adding a ``mass'' term. Such a Lagrangian is known as the de Broglie-Proca Lagrangian \citep{pro36a,deb40}, which is given by
\begin{eqnarray}
\mathcal{L}=-\frac{1}{16\pi}F_{\alpha\beta}F^{\alpha\beta}+\frac{\mu^2}{8\pi}A_\alpha A^\alpha-\frac{1}{c}J_\alpha A^\alpha,
\end{eqnarray}
where $A_\alpha=(\phi,\bm{A})$ is the gauge potential, $J_\alpha=(\rho c,\bm{J})$ is the external current sources, the field $F^{\alpha\beta}$ is given by $F^{\alpha\beta}=\partial^\alpha A^\beta-\partial^\beta A^\alpha$, and $\mu\equiv m_\gamma c/\hbar$, where $m_\gamma$ is the photon mass. The de Broglie-Proca equation reads
\begin{eqnarray}
\partial^\beta F_{\beta\alpha}+\mu^2 A_\alpha=\frac{4\pi}{c}J_\alpha.
\end{eqnarray}
In Lorenz gauge, according to current conservation, the above equation can be written as
\begin{eqnarray}
(\square+\mu^2)A_\alpha=\frac{4\pi}{c}J_\alpha.
\end{eqnarray}
Each component of $A_\alpha$ satisfies the Klein-Gordon equation with source, where
$\square\equiv\partial^2/c^2\partial t^2-\nabla^2$.
The massive photon version of Maxwell's equations is given by the de Broglie-Proca equations in three dimensions \citep{deb40}, i.e.
\begin{eqnarray}
\nabla\cdot\bm{E}&=&4\pi\rho-\mu^2\phi,\nonumber\\
\nabla\times\bm{E}&=&-\frac{1}{c}\frac{\partial \bm{B}}{\partial t},\nonumber\\
\nabla\cdot\bm{B}&=&0,\nonumber\\
\nabla\times\bm{B}&=&\frac{4\pi}{c}\bm{J}+\frac{1}{c}\frac{\partial \bm{E}}{\partial t}-\mu^2\bm{A}.
\end{eqnarray}
The associated Poynting vector is
\begin{eqnarray}
\bm{S}=\frac{c}{4\pi}(\bm{E}\times\bm{B}+\mu^2\phi\bm{A}).
\end{eqnarray}

Next, we consider the radiation of massive photons with a certain frequency. We assume that a point source of strength $f(t)$ resides at the origin. The spherical wave $\varphi(r,t)$ caused by such a source is given by
\begin{eqnarray}
(\square+\mu^2)\varphi(r,t)=\delta(\bm{r})f(t).
\end{eqnarray}
For an outgoing wave with $f(t)$ as a function of $\exp(i\omega t)$, one has \citep{cra84}
\begin{eqnarray}
\varphi(r,t)\propto\frac{1}{4\pi r}\exp\left[i\omega t-ir\left(\omega^2/c^2-\mu^2\right)^{1/2}\right].
\label{wave_equation}
\end{eqnarray}
Therefore, the dispersion relation is given by \citep{deb40}
\begin{eqnarray}
\omega^2=c^2k^2+\mu^2c^2.
\end{eqnarray}
This is the standard energy-momentum expression in the special theory of relativity.
The group velocity is variable with frequency \citep{deb40}, 
\begin{eqnarray}
\upsilon_g=c\left(1-\frac{\mu^2c^2}{\omega^2}\right)^{1/2},
\end{eqnarray}
which means that the wave is dispersed and the anomaly at $\omega=\mu c$ is representative. Since massive photons with different energies have different velocities, one can use extragalactic sources to constrain the photon mass \citep{lov64,wu16,zha16,wei16,bon16,bon17,sha17}. Due to the significant dispersion of the electromagnetic wave at very low frequencies, stringent constraints on the photon mass may be achieved by experiments at very low frequencies, e.g. the nano satellite concept \citep{ben17}.

\section{Magnetic dipole radiation with nonzero photon mass}

Following \citet{cra84},
we assume that the electric dipole moment is $\bm{p}$. In the long-wave-length limit, the vector potential is given by the integral of solutions Eq.(\ref{wave_equation}), i.e.
\begin{eqnarray}
\bm{A}(r,\theta)=\frac{i\omega \bm{p}}{cr}\exp[i(\omega t-kr)],
\end{eqnarray}
where $k=(\omega^2/c^2-\mu^2)^{1/2}$ is the dispersion relation. The magnetic and electric fields are given by
\begin{eqnarray}
\bm{B}&=&\nabla\times\bm{A}=\frac{k\omega}{cr}(\bm{n}\times\bm{p})\exp[i(\omega t-kr)],\nonumber\\
\bm{E}&=&-\frac{ic}{\omega}(\nabla\times\bm{B}+\mu^2\bm{A})=\frac{1}{cr}[\omega^2\bm{p}-k^2\bm{n}(\bm{n}\cdot\bm{p})]\exp[i(\omega t-kr)],
\end{eqnarray}
where $\bm{n}$ is the unit vector from the origin to $(r,\theta)$. The Poynting vector can be written as 
\begin{eqnarray}
\bm{P}=\frac{c}{8\pi}{\rm Re}(\bm{E}\times\bm{B^\ast}+\mu^2\phi\bm{A^\ast}).
\end{eqnarray}
The time-averaged power radiated per unit solid angle by the oscillating dipole moment $\bm{p}$ is 
\begin{eqnarray}
\frac{dL}{d\Omega}=\lim_{r\rightarrow\infty}{\rm Re}(r^2\bm{n}\cdot\bm{P})
=\frac{\omega p^2{\rm Re}(k)}{8\pi}(k^2\sin^2\theta+\mu^2). 
\label{eq:B4}
\end{eqnarray}
Therefore, the total radiation power is 
\begin{eqnarray}
L=\frac{p^2\omega}{3}\left(\frac{\omega^2}{c^2}-\mu^2\right)^{1/2}\left(\frac{\omega^2}{c^2}+\frac{\mu^2}{2}\right)
\end{eqnarray}
for $\omega>\mu c$; and $L=0$ for $\omega\leqslant\mu c$.
For the magnetic dipole field, $\bm{E}\rightarrow\bm{B}$, $\bm{B}\rightarrow-\bm{E}$, and $\bm{p}\rightarrow\bm{m}$, where $m$ is the magnetic dipole moment, we obtain the symmetric result of the total radiation power of the magnetic dipole field in vacuum as 
\begin{eqnarray}
L_m=\frac{m^2\omega}{3}\left(\frac{\omega^2}{c^2}-\mu^2\right)^{1/2}\left(\frac{\omega^2}{c^2}+\frac{\mu^2}{2}\right)
\end{eqnarray}
for $\omega>\mu c$; and $L_m=0$ for $\omega\leqslant\mu c$.

\section{Pulsar spin down power}

Here, we calculate the pulsar spindown luminosity using the method of \citet{con99}. At first, we define the enclosed magnetic flux of the open field line region as 
\begin{eqnarray}
\psi_{\rm open}&\equiv&\frac{1}{2\pi}\int \bm{B}\cdot d\bm{S}=\int_{R_{\rm LC}}^\infty B_\theta r dr\nonumber\\
&=&\psi_{\rm dipole}(\mu R_{\rm LC}+1)e^{-\mu R_{\rm LC}},\label{mf}
\end{eqnarray}
where $\psi_{\rm dipole}= B_pR^3/2R_{\rm LC}$ is the magnetic flux of the open field line in the standard magneto-static dipole model. The last equality is derived from Eq.(\ref{mag}).
We assume that the flux distribution along the open field lines is close to the Michel split-monopole solution \citep{mic74}, e.g.,
\begin{eqnarray}
I(\psi)\simeq I_{\rm Michel}=\psi\left(2-\frac{\psi}{\psi_{\rm open}}\right).
\end{eqnarray}
Due to magnetospheric rotation, the electric current circuits are generated at the pulsar poles, forming electromagnetic torques anti-parallel to the angular momentum of the pulsar, e.g., $T=rBJdSdr/c$, where $dS$ denotes any stellar cross section, and $J$ denotes the poloidal electric current density $J$.
Finally, the stellar rotation energy loss through the electromagnetic torques is given by \citep{con05}
\begin{eqnarray}
L_w=\frac{\Omega^2}{c}\int_{\psi=0}^{\psi_{\rm open}} I(\psi)d\psi=\frac{2}{3}\frac{\Omega^2}{c}\psi_{\rm open}^2=\eta L_{w,0},
\end{eqnarray}
where $L_{w,0}=(2/3)\Omega^2\psi_{\rm dipole}^2/c$ is the classical magnetic dipole radiation power, and 
\begin{eqnarray}
\eta=\left(1+\frac{\mu c}{\Omega}\right)^2e^{-2\mu c/\Omega}.
\end{eqnarray}
This result is consistent with Eq.(\ref{wp}).

\end{document}